\documentclass[aps,showpacs,preprintnumbers,amsmath, amssymb]{revtex4}

\usepackage{float}
\usepackage{graphics,epsfig}
\usepackage{graphicx}
\usepackage{dcolumn}
\usepackage{bm}

\begin{document}
\baselineskip=0.8 cm
\title{{\bf Quasinormal modes and late-time tails in the background of Schwarzschild black hole pierced
 by a cosmic string: scalar, electromagnetic and gravitational perturbations}}
\author{Songbai Chen}
\email{chsb@fudan.edu.cn} \affiliation{Department of Physics,
Fudan University, Shanghai 200433, P. R. China
 \\ Institute of Physics and  Department of Physics,
Hunan Normal University,  Changsha, Hunan 410081, P. R. China }

\author{Bin Wang}
\email{wangb@fudan.edu.cn} \affiliation{Department of Physics,
Fudan University, Shanghai 200433, P. R. China}

\author{ Rukeng Su}
\email{rksu@fudan.ac.cn}
 \affiliation{China Center of Advanced Science and Technology (World Laboratory),
P.B.Box 8730, Beijing 100080, People¡¯s Republic of China
\\ Department of Physics, Fudan University, Shanghai 200433, P. R. China}

\vspace*{0.2cm}
\begin{abstract}
\baselineskip=0.6 cm
\begin{center}
{\bf Abstract}
\end{center}

We have studied the quasinormal modes and the late-time tail
behaviors of scalar, electromagnetic and gravitational
perturbations in the Schwarzschild black hole pierced by a cosmic
string. Although the metric is locally identical to that of the
Schwarzschild black hole so that the presence of the string will
not imprint in the motion of test particles, we found that
quasinormal modes and the late-time tails can reflect physical
signatures of the cosmic string. Compared with the scalar and
electromagnetic fields, the gravitational perturbation decays
slower, which could be more interesting to disclose the string
effect in this background.

\end{abstract}

\pacs{ 04.30.Nk, 04.70.Bw} \maketitle
\newpage
\vspace*{0.2cm}
\section{Introduction}

Quasinormal mode (QNM) is believed as a characteristic sound of
black holes, which describes the damped oscillations under
perturbations in the surrounding geometry of a black hole with
frequencies and damping times of the oscillations entirely fixed
by the black hole parameters. The study of QNM has been an
intriguing subject of discussions for the last few decades
\cite{1,2,3}. Despite its potential astrophysical interest which
could lead to the direct identification of the black hole
existence through gravitational wave observation to be realized in
the near future\cite{1,2}, the QNM has also been argued as a
testing ground for fundamental physics. Motivated by the discovery
of the AdS/CFT correspondence, the investigation of QNM in anti-de
Sitter(AdS) spacetimes became appealing in the past several years.
It was argued that the QNMs of AdS black holes have direct
interpretation in term of the dual conformal field
theory(CFT)\cite{3,4,5,6,7,8,9,13}. Attempts of using QNMs to
investigate the dS/CFT correspondence has also been
given\cite{10}. Recently QNMs in asymptotically flat spaces have
acquired further attention, since the possible connection between
the classical vibrations of a black hole spacetime and various
quantum aspects was proposed by relating the real part of the QNM
frequencies to the Barbero-Immirzi(BI) parameter, a factor
introduced by hand in order that loop quantum gravity reproduces
correctly the black hole entropy \cite{11}. The extension has been
done in the dS background \cite{12}, however in the AdS black hole
spacetime, the direct relation has not been found\cite{13}. The
charged situation in both dS and AdS cases were fully addressed in
\cite{aa}.

The motivation of the present paper is to examine whether the QNM
can tell us the signature of the string. We are going to consider
curved spacetimes containing strings. An example is a black hole
with a straight string passing through it, whose metric is
described as \cite{14}\cite{15}
\begin{eqnarray}
ds^2=(1-\frac{2M}{br})dt^2-(1-\frac{2M}{br})^{-1}dr^2
-r^2d\theta^2-b^2r^2\sin^2{\theta}d\varphi^2,\label{metric}
\end{eqnarray}
where $M$ is the parameter which relates to the black hole energy.
$b$ is related to the linear mass density $\rho_s$ of the string
by $b=1-4\rho_s$ and is justified in the range $0<b\leq 1$ since
$\rho_s\ll 1$. The metric is locally identical to the
Schwarzschild metric, and the motion of test particles is locally
unchanged by the presence of the string. Thus it would be
difficult to observe the string influence through its effect on
the geodesics. It is of interest to ask what physical phenomenon
will present due to the presence of the string. In this work we
will examine the QNM in the background of Schwarzschild black hole
pierced by a cosmic string and will disclose phenomenons in the
QNM in terms of the cosmic string.

We will extend the study of the wave dynamics outside the black
hole to the late time scale. At late times, quasinormal
oscillations are swamped by the relaxation process. This
relaxation is the requirement of the black hole no-hair theorem
\cite{Wheeler1}. Price \cite{Price1} first studied the late-time
behaviors of the massless scalar, gravitational and
electromagnetic perturbations in the Schwarzschild black hole
spacetime and found that for a fixed $r$ the late-time tails are
commonly dominated by the factor $t^{-(2l+3)}$. This late-time
behavior has also been confirmed in other static spacetimes for
massless scalar perturbations \cite{Hod2}. For the massive scalar
field, the late-time behaviors of perturbations persist quite
different properties. For example, the intermediate late-time
behavior is dominated by the oscillatory inverse power-law form
$t^{-(l+3/2)}\sin{\mu t}$ and the asymptotic one is by the
$t^{-5/6}\sin{\mu t}$ in the static spacetimes \cite{Koyama}.
Similar properties of the late-time behaviors have also been found
in the rotating black hole backgrounds \cite{Hod3}. For a black
hole with a straight string passing through it, the black hole may
have hairs in the form of strings\cite{15}. It would be
interesting to investigate whether the late-time tail contains the
information of the string.

The organization of this paper is as follows. In Sec.II we will
derive the wave equation of the scalar, electromagnetic and
gravitational perturbations in the Schwarzschild black hole
pierced by a cosmic string. In Sec.III we will calculate the
fundamental quasinormal frequencies of the massless scalar,
electromagnetic and gravitational fields by using the third-order
WKB approximation. In Sec.IV we will extend our discussion to the
late-time tail behaviors of the massless scalar, electromagnetic
and gravitational perturbations. For the massive late-time tail
behaviors, we just consider the scalar perturbation. Our
conclusions and discussions will be presented in the last section.

\section{The wave equation of the scalar,  electromagnetic and
gravitational perturbations in the Schwarzschild black hole
pierced by a cosmic string}

For the Schwarzschild black hole pierced by a cosmic string, we
can choose the the null tetrad as follows
\begin{eqnarray}
l^{\mu}&=&\bigg(\frac{r^2}{\Delta_r},1,0,0 \bigg),\nonumber\\
n^{\mu}&=&\frac{1}{2}\bigg(1,-\frac{\Delta_r}{r^2},0,0 \bigg),\nonumber\\
m^{\mu}&=&\frac{1}{\sqrt{2}r}\bigg(0,0,1,\frac{i}{b\sin{\theta}} \bigg),\nonumber\\
\end{eqnarray}
where $\Delta_r=r(r-2M/b)$, and we find that the non-vanishing
spin-coefficients have the values
\begin{eqnarray}
\rho=-\frac{1}{r},\;\;\;\;\;
\beta=-\alpha=\frac{1}{2\sqrt{2}r}\cot{\theta},\;\;\;\;\;
\mu=\frac{\Delta_r}{2r^3},\;\;\;\;\;\gamma=\frac{M}{2br^2}.
\end{eqnarray}
The Teukolsky's master equations \cite{21} for massless scalar
($s=0$), electromagnetic ($s=1$) and gravitational ($s=2$)
perturbations in Newman-Penrose formalism can be written as
\cite{22}
\begin{eqnarray}
&&\{[D-(2s-1)\varepsilon+\varepsilon^*-2s\rho-\rho^*](\Delta-2s\gamma+\mu)\nonumber\\
&&-[\delta-(2s-1)\beta-\alpha^*-2s\tau+\pi^*](\bar{\delta}-2s\alpha+\pi)-(s-1)(2s-1)\Psi_2\}\Phi_s=0,
\label{E4}
\end{eqnarray}
and
\begin{eqnarray}
&&\{[\Delta+(2s-1)\gamma-\gamma^*+2s\mu+\mu^*](D+2s\varepsilon-\rho)\nonumber\\
&&-[\bar{\delta}+(2s-1)\alpha+\beta^*+2s\pi-\tau^*](\delta+2s\beta-\tau)-(s-1)
(2s-1)\Psi_2\}\Phi_{-s}=0.\label{E5}
\end{eqnarray}
Assuming that the azimuthal and time dependence of the
wave-functions in equations (\ref{E4}) and (\ref{E5}) have the
form $e^{-i(\omega t-m\varphi)}$, we find that the derivative
operators are
\begin{eqnarray}
&&D\equiv l^\mu \partial_{\mu}=\mathcal{D}_0,\;\;\;\;\;\;\;\;\;\;
\Delta\equiv n^{\mu}\partial_{\mu}=-\frac{\Delta_r}{2r^2}\mathcal{D}^{\dag}_0,\nonumber\\
&&\delta\equiv
m^{\mu}\partial_{\mu}=\frac{1}{\sqrt{2}r}\mathcal{L}^{\dag}_0,
\;\;\;\;\; \bar{\delta}\equiv
m^{*\mu}\partial_{\mu}=\frac{1}{\sqrt{2}r}\mathcal{L}_0,
\end{eqnarray}
where
\begin{eqnarray}
&& \mathcal{D}_n=\frac{\partial}{\partial r}-\frac{i
K}{\Delta_r}+\frac{n}{\Delta_r}\frac{d\Delta_r}{dr},\;\;\;\;\;\;\;\;\;\;
\mathcal{D}^{\dag}_n=\frac{\partial}{\partial r}+\frac{i
K}{\Delta_r}+\frac{n}{\Delta_r}\frac{d\Delta_r}{dr},\nonumber\\
&& \mathcal{L}_n=\frac{\partial}{\partial \theta}
+\frac{m}{b\sin{\theta}}+n\cot{\theta},\;\;\;\;\;\;\;\;\;\;
\mathcal{L}^{\dag}_n=\frac{\partial}{\partial \theta}
-\frac{m}{b\sin{\theta}}+n\cot{\theta},\;\;\;\;\;\; K=r^2\omega.
\end{eqnarray}
Adopting the Newman-Penrose formalism, we can easily obtain the
separated equations\cite{22} for massless scalar, electromagnetic
and gravitational perturbations around a Schwarzschild black hole
pierced by a cosmic string
\begin{eqnarray}
&& [\Delta_r\mathcal{D}_{1-s}\mathcal{D}^{\dag}_0+2(2s-1)i\omega
r]\Delta^s_rR_s=\lambda\Delta^s_rR_s,\nonumber\\
&& \mathcal{L}^{\dag}_{1-s}\mathcal{L}_sS_s=-\lambda
S_s,\label{E8}
\end{eqnarray}
with the spin number $s = 0, +1$ and $+2$, where $\lambda$ is  a
separation constant, $R_s$ and $S_s$ are only functions of $r$ and
$\theta$, respectively.

Using the transformation theory, we can obtain the radial equation
for the massless scalar, electromagnetic and gravitational
perturbations in this black hole spacetime through tedious
calculations, which leads to the form
\begin{eqnarray}
\frac{d^2 P_s}{dr^2_*}+[\omega^2-V_s]P_s=0,
\end{eqnarray}
where $r_*$ is the tortoise coordinate (which is defined by
$dr_*=\frac{r^2}{\Delta_r}dr$) and the effective potential $V_s$
reads
\begin{eqnarray}
V_s=\left\{
\begin{array}{l}(1-\frac{2M}{br})(\frac{\lambda}{r^2}+\frac{2M}{br^3})
\;\;\;\;\;\;\;\;\; s=0,\\ \\
(1-\frac{2M}{br})\frac{\lambda}{r^2}\;
\;\;\;\;\;\;\;\;\;\;\;\;\;\;\;\;\;\;\;\; \; s=1,\\
\\
(1-\frac{2M}{br})(\frac{\lambda+2}{r^2}-\frac{6M}{br^3}),
\;\;\;\;\;\; s=2.
\end{array}\right.\label{efp}
\end{eqnarray}
From the second equation in (\ref{E8}), we can obtain directly the
angular equation for the perturbations
\begin{eqnarray}
\bigg[\frac{1}{\sin{\theta}}\frac{d}{d\theta}\bigg(\sin{\theta}\frac{d}{d\theta}\bigg)
-\frac{(m/b+s\cos{\theta})^2}{\sin^2{\theta}}+s\bigg] S_s=-\lambda
S_s,\label{angd}
\end{eqnarray}
where $m$ is the angular quantum number. For the parameter $b=1$
(i.e., the linear mass density of the cosmic string $\rho_s=0$),
the angular equation (\ref{angd}) reduces to that in the usual
spherically symmetric case. The solution can be expressed as the
expansion in the spin-weighted associated Legendre polynomial
$_sP_{lm}(\cos{\theta})$ \cite{23} with the eigenvalue
$\lambda=(l+s)(l-s+1)$ which is independent of the angular number
$m$. However when $b\neq 1$, the spherical symmetry is broken and
then $\lambda$ will depend not only on the quantum number $l$ but
also on $m$.

To investigate the quasinormal modes and late-time tail behaviors
of the external perturbations in the black hole spacetime, we must
first determine $\lambda$ appeared in the above equations. We
restrict ourselves to $m>0$ and rewrite equation (\ref{angd}) as
\begin{eqnarray}
\bigg[\frac{1}{\sin{\theta}}\frac{d}{d\theta}\bigg(\sin{\theta}\frac{d}{d\theta}\bigg)
-\frac{(m+s\cos{\theta})^2}{\sin^2{\theta}}+s+\frac{(b^2-1)m^2}{b^2\sin^2{\theta}
}+\frac{2ms(b-1)\cos{\theta}}{b \sin^2{\theta}}\bigg] S_s=-\lambda
S_s. \label{angd1}
\end{eqnarray}
Considering that the deviation of parameter $b$ from the unity is
very small, which is physically justified for a cosmic string with
$\rho_s\ll 1$, the last two terms in the left-hand-side of the
above equation can be regarded as a perturbative term. Using the
general perturbation theory, we have
\begin{eqnarray}
S_s&=&_sP_{lm}(\cos{\theta})+\gamma\; _s\zeta_{lm}(\cos{\theta})+O(\gamma^2),\nonumber\\
\lambda&=&\lambda_0+\gamma\lambda_1+O(\gamma^2),\label{pe1}
\end{eqnarray}
where $\gamma$ is a dimensionless parameter denoting the
perturbative scale. For convenience, we set $\gamma=1$
throughout our paper. Substituting the variables (\ref{pe1})
into the angular equation (\ref{angd1}), we can obtain
\begin{eqnarray}
&&(D_0+\lambda_0)\;_sP_{lm}(\cos{\theta})=0,\label{zero}\\
&&(D_0+\lambda_0)\;
_s\zeta_{lm}(\cos{\theta})+(D_1+\lambda_1)\;_sP_{lm}(\cos{\theta})=0,\label{1-order}
\end{eqnarray}
where
\begin{eqnarray}
&&D_0=\frac{1}{\sin{\theta}}\frac{d}{d\theta}\bigg(\sin{\theta}\frac{d}{d\theta}\bigg)
-\frac{(m/b+s\cos{\theta})^2}{\sin^2{\theta}}+s,\nonumber\\
&&D_1=\frac{(b^2-1)m^2}{b^2\sin^2{\theta}
}+\frac{2ms(b-1)\cos{\theta}}{b \sin^2{\theta}}.
\end{eqnarray}
From the zeroth order equation (\ref{zero}), we have
\begin{eqnarray}
\lambda_0=(l+s)(l-s+1).
\end{eqnarray}
Multiplying equation (\ref{1-order}) by $_sP_{lm}(\cos{\theta})$
from the left side and integrating it over $\theta$, we obtain
\begin{eqnarray}
\lambda_1=\left\{
\begin{array}{l} \frac{m(2l+1)(1-b)}{2(m^2-s^2)b^2}[(1+b)m^2-2bs^2]
\;\;\;\;\;\;\;\;\; |m|>|s|,\\ \\
\frac{sm^2(2l+1)(1-b)^2}{2(s^2-m^2)b^2}
\;\;\;\;\;\;\;\;\;\;\;\;\;\;\;\;\;\;\;\; \;\;\;\;\;\;\;\;\;\;
\;\;\;\; |m|<|s|.
\end{array}\right.\label{efp1}
\end{eqnarray}
As we discussed above, when $b\neq 1$ the eigenvalue $\lambda$ of
the angular equation (\ref{angd}) depends not only on the multiple
moment $l$, but also on the parameter $b$, angular number $m$ and
the spin $s$. Obviously, the eigenvalue $\lambda$ increases with
the increase of the parameters $m$ and $\rho_s$. At last, it must
to be noted that the equation (\ref{efp1}) is not valid as
$|m|=|s|$.

\section{Quasinormal modes of scalar, electromagnetic and
gravitational perturbations in the Schwarzschild black hole
pierced by a cosmic string}

In this section, we will apply the third-order WKB approximation
to evaluate the fundamental quasinormal modes ($n=0$) of massless
scalar, electromagnetic and gravitational perturbations in the
Schwarzschild black hole with a cosmic string passing through it.
We expect to see what effects of cosmic string can be reflected in
the QNMs behavior. The formula for the complex quasinormal
frequencies $\omega$ in this approximation is given by \cite{wkb}
\begin{eqnarray}
\omega^2=[V_0+(-2V^{''}_0)^{1/2}\Lambda]-i(n+\frac{1}{2})(-2V^{''}_0)^{1/2}(1+\Omega),
\end{eqnarray}
where
\begin{eqnarray}
\Lambda&=&\frac{1}{(-2V^{''}_0)^{1/2}}\left\{\frac{1}{8}\left(\frac{V^{(4)}_0}{V^{''}_0}\right)
\left(\frac{1}{4}+\alpha^2\right)-\frac{1}{288}\left(\frac{V^{'''}_0}{V^{''}_0}\right)^2
(7+60\alpha^2)\right\},\nonumber\\
\Omega&=&\frac{1}{(-2V^{''}_0)}\bigg\{\frac{5}{6912}
\left(\frac{V^{'''}_0}{V^{''}_0}\right)^4
(77+188\alpha^2)\nonumber\\&-&
\frac{1}{384}\left(\frac{V^{'''^2}_0V^{(4)}_0}{V^{''^3}_0}\right)
(51+100\alpha^2)
+\frac{1}{2304}\left(\frac{V^{(4)}_0}{V^{''}_0}\right)^2(67+68\alpha^2)
\nonumber\\&+&\frac{1}{288}
\left(\frac{V^{'''}_0V^{(5)}_0}{V^{''^2}_0}\right)(19+28\alpha^2)-\frac{1}{288}
\left(\frac{V^{(6)}_0}{V^{''}_0}\right)(5+4\alpha^2)\bigg\},
\end{eqnarray}
and
\begin{eqnarray}
\alpha=n+\frac{1}{2},\;\;\;\;\;
V^{(s)}_0=\frac{d^sV}{dr^s_*}\bigg|_{\;r_*=r_*(r_{p})} \nonumber,
\end{eqnarray}
$n$ is overtone number and $r_{p}$ is the value of polar
coordinate $r$ corresponding to the peak of the effective
potential (\ref{efp}).

Setting $M=1$ and substituting the effective potential (\ref{efp})
into the formula above, we can obtain the quasinormal frequencies
of scalar, electromagnetic and gravitational perturbations in the
Schwarzschild black hole pierced by a cosmic string.
\begin{table}[h]
\begin{center}
\begin{tabular}[b]{cccc}
 \hline \hline
 \;\;\;\; $b$ \;\;\;\; & \;\;\;\; $\omega\ \ \ (m=0)$\;\;\;\;  & \;\;\;\;  $\omega \ \ \ (m=1)$\;\;\;\;
 & \;\;\;\; $\omega \ \ \ (m=2)$ \;\;\;\; \\ \hline
\\
0.90& \;\;\;\;\;0.953613-0.086705i\;\;\;\;\;  & \;\;\;\;
0.973696-0.086701i\;\;\;\;\;
 & \;\;\;\;\;0.993374-0.086697i\;\;\;\;\;
 \\
0.92&0.974805-0.088632i&0.990725-0.088628i&1.006393-0.088625i
 \\
0.94&0.995996-0.090559i&1.007830-0.090556i&1.019526-0.090554i
 \\
0.96&1.017188-0.092485i&1.025008-0.092484i&1.032770-0.092482i
\\
0.98&1.038379-0.094412i&1.042256-0.094411i&1.046119-0.094410i
\\
1.00&1.059570-0.096339i&1.059570-0.096339i&1.059570-0.096339i
\\
 \hline \hline
 $b$  &$\omega \ \ \ (m=3)$ & $\omega \ \ \ (m=4)$ & $\omega \ \ \ (m=5)$\\ \hline
 \\
0.90&1.012668-0.086693i&1.031602-0.086690i&1.050194-0.086687i
\\
0.92&1.021820-0.088622i&1.037019-0.088619i&1.051997-0.088617i
\\
0.94&1.031090-0.090551i&1.042525-0.090549i&1.053837-0.090547i
\\
0.96&1.040473-0.092480i&1.048120-0.092479i&1.055712-0.092477i
\\
0.98&1.049968-0.094410i&1.053803-0.094409i&1.057623-0.094408i
\\
1.00&1.059570-0.096339i&1.059570-0.096339i&1.059570-0.096339i
\\ \hline\hline
\end{tabular}
\end{center}
\caption{The fundamental ($n=0$) quasinormal frequencies of scalar
field (s=0) in the Schwarzschild black hole spacetime with a
cosmic string for $l=5$.}
\end{table}
\begin{table}[h]
\begin{center}
\begin{tabular}[b]{cccc}
 \hline \hline
 \;\;\;\; $b$ \;\;\;\; & \;\;\;\; $\omega\ \ \ (m=0)$\;\;\;\;  & \;\;\;\;  $\omega \ \ \ (m=2)$\;\;\;\;
 & \;\;\;\; $\omega \ \ \ (m=3)$ \;\;\;\; \\ \hline
\\
0.90& \;\;\;\;\;0.943084-0.086386i\;\;\;\;\;  & \;\;\;\;
0.983959-0.086403i\;\;\;\;\;
 & \;\;\;\;\;1.003138-0.086410i\;\;\;\;\;
 \\
0.92&0.964041-0.088305i&0.996406-0.088319i&1.011798-0.088325i
 \\
0.94&0.984999-0.090225i&1.009027-0.090235i&1.020604-0.090240i
 \\
0.96&1.005956-0.092145i&1.021815-0.092152i&1.029555-0.092155i
\\
0.98&1.026913-0.094064i&1.034765-0.094068i&1.038645-0.094069i
\\
1.00&1.047871-0.095984i&1.047871-0.095984i&1.047871-0.095984i
\\
 \hline \hline
 $b$  &$\omega \ \ \ (m=4)$ & $\omega \ \ \ (m=5)$\\ \hline
 \\
0.90&1.022140-0.086417i&1.040844-0.086424i
\\
0.92&1.027075-0.088331i&1.042160-0.088337i
\\
0.94&1.032117-0.090245i&1.043520-0.090249i
\\
0.96&1.037265-0.092158i&1.044926-0.092161i
\\
0.98&1.042517-0.094071i&1.046376-0.094073i
\\
1.00&1.047871-0.095984i&1.047871-0.095984i
\\ \hline\hline
\end{tabular}
\end{center}
\caption{The fundamental ($n=0$) quasinormal frequencies of
electromagnetic field (s=1) in the Schwarzschild black hole
spacetime with a cosmic string for $l=5$.}
\end{table}

\begin{table}[h]
\begin{center}
\begin{tabular}[b]{cccc}
 \hline \hline
 \;\;\;\; $b$ \;\;\;\; & \;\;\;\; $\omega\ \ \ (m=0)$\;\;\;\;  & \;\;\;\;  $\omega \ \ \ (m=1)$\;\;\;\;
 & \;\;\;\; $\omega \ \ \ (m=3)$ \;\;\;\; \\ \hline
\\
0.90& \;\;\;\;\;0.911027-0.085386i\;\;\;\;\;  & \;\;\;\;
0.911771-0.085388i\;\;\;\;\;
 & \;\;\;\;\;0.975143-0.085534i\;\;\;\;\;
 \\
0.92&0.931272-0.087283i&0.931738-0.087285i&0.981941-0.087403i
 \\
0.94&0.951517-0.089181i&0.951774-0.089182i&0.989058-0.089271i
 \\
0.96&0.971762-0.091078i&0.971874-0.091079i&0.996487-0.091139i
\\
0.98&0.992007-0.092976i&0.992034-0.092976i&1.004220-0.093006i
\\
1.00&1.012252-0.094873i&1.012252-0.094873i&1.012252-0.094873i
\\
 \hline \hline
 $b$  &$\omega \ \ \ (m=4)$ & $\omega \ \ \ (m=5)$\\ \hline
 \\
0.90&0.993689-0.085571i&1.012586-0.085608i
\\
0.92&0.997042-0.087435i&1.012359-0.087466i
\\
0.94&1.000576-0.089297i&1.012212-0.089322i
\\
0.96&1.004291-0.091157i&1.012145-0.091175i
\\
0.98&1.008183-0.093016i&1.012158-0.093025i
\\
1.00&1.012252-0.094873i&1.012252-0.094873i
\\ \hline\hline
\end{tabular}
\end{center}
\caption{The fundamental ($n=0$) quasinormal frequencies of
gravitational field (s=2) in the Schwarzschild black hole
spacetime with a cosmic string for $l=5$.}
\end{table}

\begin{figure}[ht]
\begin{center}
\includegraphics[width=17cm]{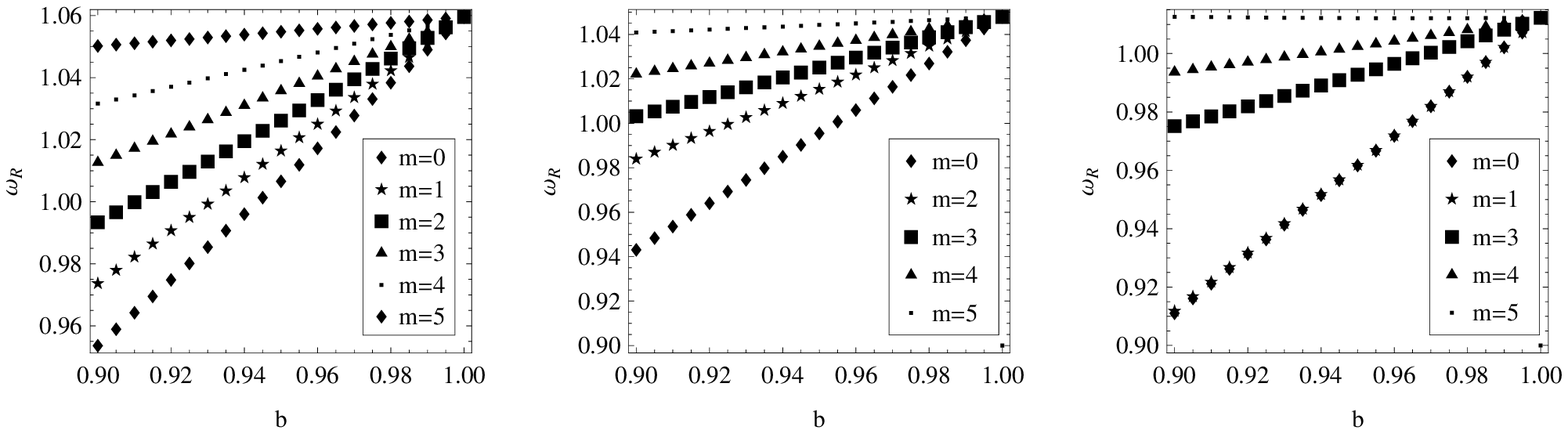}
\caption{Dependence of the real part of quasinormal frequencies
($n=0$) for scalar (the left), electromagnetic (the middle) and
gravitational (the right) perturbations on the parameter $b$ for
fixed $l=5$ and chosen $m$. }
 \end{center}
 \label{fig1}
 \end{figure}

\begin{figure}[ht]
\begin{center}
\includegraphics[width=17cm]{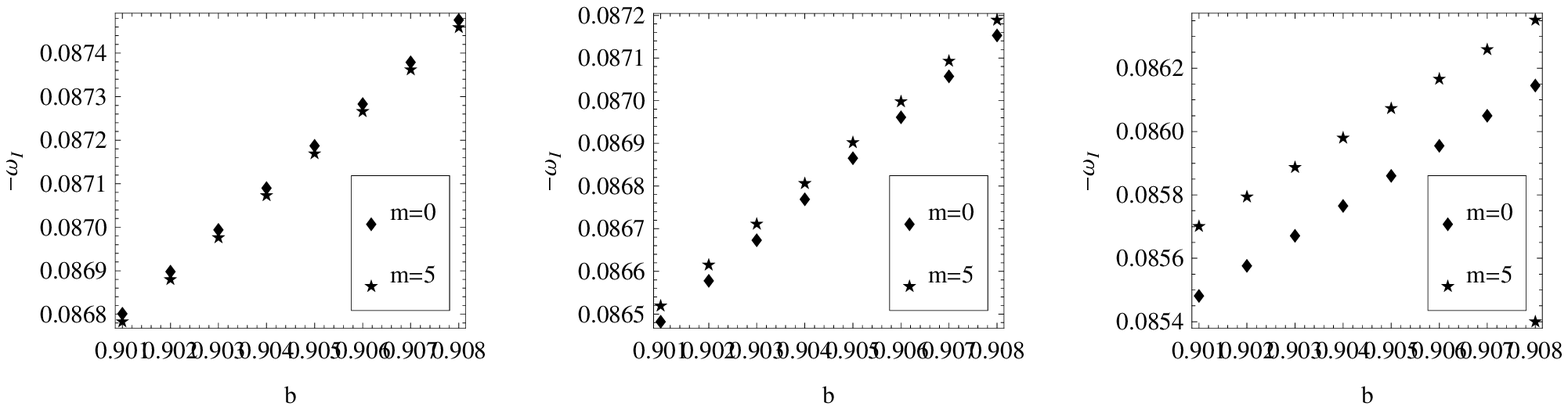}
\caption{Dependence of the imaginary part of quasinormal
frequencies ($n=0$) for scalar (the left), electromagnetic (the
middle) and gravitational (the right) perturbations on the
parameter $b$ for fixed $l=5$ and chosen $m$. }
 \end{center}
 \label{fig2}
 \end{figure}

Since the properties of the fundamental QNM frequencies for
different $l$ are similar in this background, we only list the QNM
frequencies for $l=5$ in Table 1. From figures (1), (2) and (3),
we clearly see that quasinormal frequencies for different spin $s$
fields depend not only on the parameter $b$ but also on the
angular number $m$, which is quite different from the QNMs
disclosed in the general spherically symmetric backgrounds. For
the fixed $l$, $m$ and $s$, both the real part and the absolute
value of the imaginary part of quasinormal frequencies are almost
linear functions of the parameter $b$. This is not surprising
because in terms of the P\"{o}schl-Teller potential approximation
it was found that the real and imaginary parts of quasinormal
frequencies of the scalar, electromagnetic and gravitational
fields for large multiple number $l$ in the black hole with a
cosmic string can be well approximated as $\omega_{R} \sim
\frac{b}{3\sqrt{3}M}\sqrt{l(l+1)+\lambda_1}$ and $\omega_{I} \sim
-i\frac{b}{3\sqrt{3}M}(n+\frac{1}{2})$. We have also examined the
QNMs dependence on the angular number $m$ for fixed $l$ and chosen
$b$. Results are shown in Fig.3, where we observed that with the
increase of $m$, both the real parts and absolute value of the
imaginary parts of the QNM frequencies for electromagnetic and
gravitational fields increase, while for scalar perturbation, the
real part increases and the absolute value of the imaginary part
decreases. The dependence on the angular quantum number $m$ has
not been observed for spherical cases before, it is similar to
those in the Kerr black hole spacetime \cite{Cardoso2}. Compared
with the scalar and electromagnetic perturbations, the
gravitational perturbation has both smaller real part and smaller
absolute value of the imaginary part of quasinormal frequencies.
The gravitational perturbation can thus last longer than other
perturbations, thus gravitational perturbation could be of more
interesting to detect the string effect in this background.
\begin{figure}[ht]
\begin{center}
\includegraphics[width=17cm]{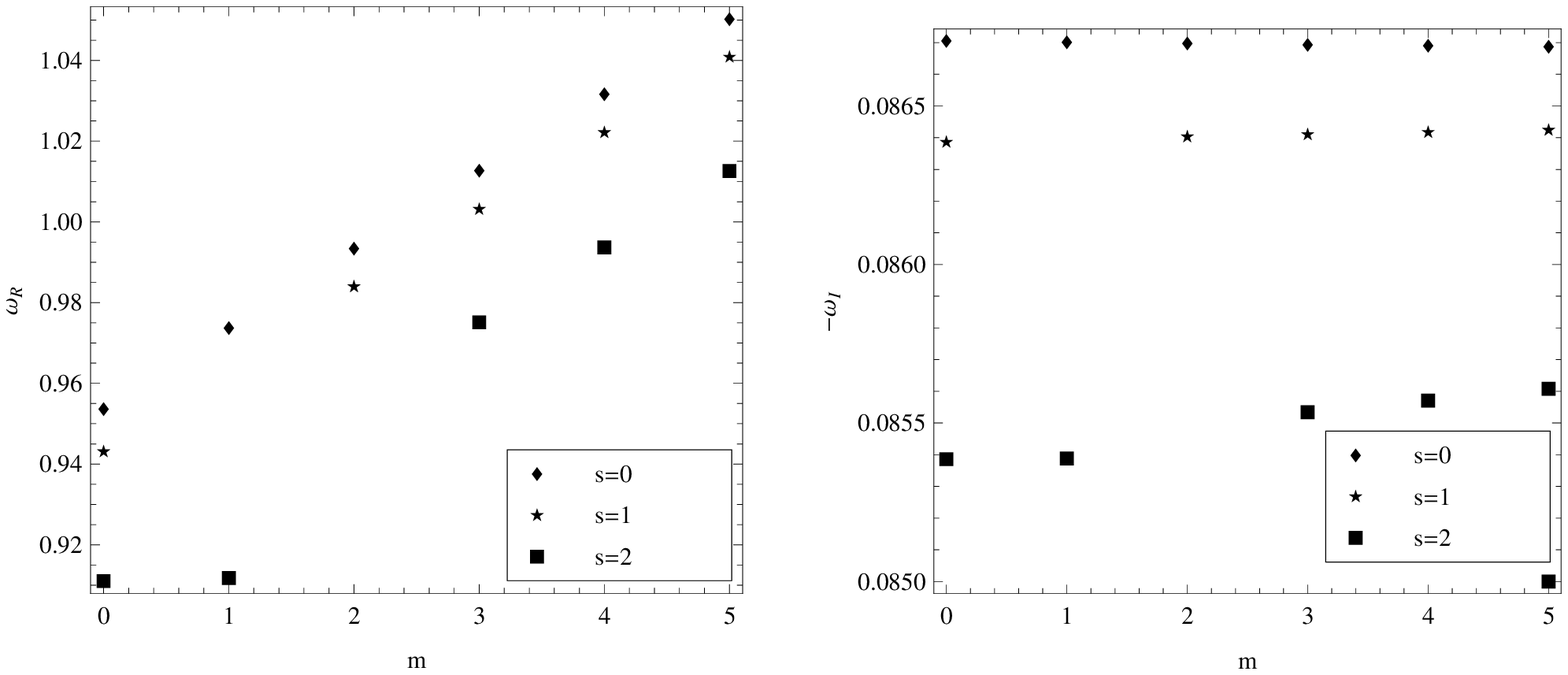}
\caption{Dependence of the real part (the left) and the imaginary
part (the right) of quasinormal frequencies ($n=0$) on the angular
number $m$ for fixed $l=5$ and $b=0.9$. }
 \end{center}
 \label{fig3}
 \end{figure}

\section{Late-time tails of scalar, electromagnetic and gravitational perturbations
in the Schwarzschild black hole pierced by a cosmic string}

In this section, we will adopt the technique of spectral
decomposition \cite{Leaver} to investigate the late-time tails of
scalar, electromagnetic and gravitational fields in the background
of the Schwarzschild black hole with a cosmic string passing
through it.

In terms of this method, the retarded Green's function
$G(r_*,r'_*;t)$, which determines the time evolution of a scalar
field in the background spacetime, can be expressed as
\begin{eqnarray}
G(r_*,r'_*;t)=\frac{1}{2\pi}\int^{\infty+ic}_{-\infty+ic}{\tilde{G}(r_*,r'_*;\omega)e^{-i\omega
t}d\omega},\label{G1}
\end{eqnarray}
where $c$ is a positive constant and $\tilde{G}(r_*,r'_*;\omega)$
is defined by
\begin{eqnarray}
\tilde{G}(r_*,r'_*;\omega)=-\frac{1}{W(\omega)}\{
\begin{array}{c}
 \tilde{\Psi}_1(r_*,\omega)\tilde{\Psi}_2(r'_*,\omega),\ \ \ r_*<r'_*;\\
\tilde{\Psi}_1(r'_*,\omega)\tilde{\Psi}_2(r_*,\omega),\ \ \
r_*>r'_*,
\end{array}
\end{eqnarray}
with
\begin{eqnarray}
W(\omega)=W(\tilde{\Psi}_1,\tilde{\Psi}_2)=\tilde{\Psi}_1\tilde{\Psi}_{2,r_*}
-\tilde{\Psi}_2\tilde{\Psi}_{1,r_*},
\end{eqnarray}
$\tilde{\Psi}_1(r_*,\omega)$ and $\tilde{\Psi}_2(r_*,\omega)$
 are linearly independent solutions to the homogeneous
equation
\begin{eqnarray}
\left[\frac{d^2}{dr^2_*}+\omega^2-V(r)\right]\tilde{\Psi}_i(r_*,\omega)=0,\
\ \ i=1,2.\label{e2}
\end{eqnarray}
It is by now well known that there exists a branch cut in
$\tilde{\Psi}_2$ placed along the negative imaginary $\omega$-axis
and the contribution of the Green function $G(r_*,r'_*; t)$ to
late-time tail comes from the integral of $\tilde{G}(r_*,r'_*;
\omega)$ around this branch cut which is denoted by $G^C(r_*,r'_*;
t)$. Thus, in the study of late-time evolution of an external
field, we just need to consider $G^C(r_*,r'_*; t)$.

\subsection{Late-time behavior of the massless perturbations}

Following Ref.\cite{Hod2}, we can adopt the low-frequency
approximation to study the asymptotic late-time behavior of the
massless perturbational fields in the Schwarzschild black hole
pierced by a cosmic string. Neglecting terms of the order
$O(\frac{M^2}{r^2} )$ and the higher order terms, we can expend
the wave equation (\ref{e2}) for the massless perturbational
fields  as a power series in $M/r$
\begin{eqnarray}
\left[\frac{d^2}{dr^2}+\omega^2+\frac{4M'\omega^2}{r}-
\frac{\rho^2-\frac{1}{4}}{r^2}\right]\zeta(r,\omega)=0,\label{19}
\end{eqnarray}
where $\zeta(r,\omega)=\sqrt{1-\frac{2M}{br}}\tilde{\Psi}$,
$M'=M/b$ and $\rho=\sqrt{(l+\frac{1}{2})^2+\lambda_1}$. We can
directly obtain that the asymptotic behavior of the massless
scalar, electromagnetic and gravitational fields at timelike
infinity in the background of the Schwarzschild black hole pierced
by a cosmic string
\begin{eqnarray}
G^C(r_*,r'_*;t)&=&\frac{2^{2\rho}M[\Gamma{(\rho+\frac{1}{2})}]^2(-1)^{2\rho+1}\Gamma{(2\rho+2)}}{\pi
b [\Gamma{(2\rho+1)}]^2}(r'_*r_*)^{\rho+1/2}
t^{-2\rho-2}.\label{mlg}
\end{eqnarray}
It is easy to find that different from the usual spherical black
hole, the late-time tails of the massless scalar, electromagnetic
and gravitational fields in this background are related not only
to the multiple moment $l$, but also to the parameter $b$, which
has the connection to the linear mass energy of the cosmic string,
the angular number $m$ and the spin $s$ of perturbational fields.
From figure 4, we can obtain that for the fixed $m$, the larger
the parameter $b$ is, the more slowly the perturbation decays.
Moreover, figure 5 also tells us that for the fixed $b$, the
larger the parameter $m$ is, the more quickly the perturbation
dies out. From figure 6, we also find that the larger the spin $s$
is, the more slowly the perturbation decays, which means the
gravitational field decays slower, which is consistent with the
discussion above in the QNM investigation. These results are
different from that in the usual Schwarzschild case with $b=1$,
where the damping exponent of the perturbational fields are
independent of $m$ and $s$.
\begin{figure}[H]
\begin{center}
\includegraphics[width=17cm]{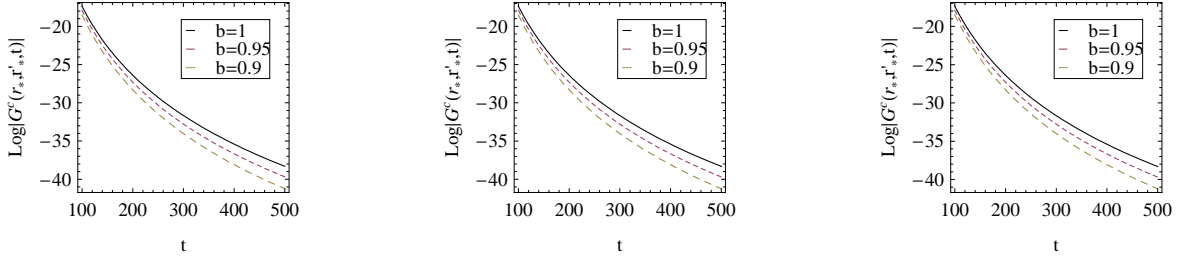}\caption{
Graphs of $\ln G^C(r_*, r'_*; t)$ (the left $s=0$, the middle
$s=1$, the right $s=2$) versus $t$ for the change of $b$, with
$l=5$ and $m=5$. Here, $M=1$, $r_*=10$ and $r'_*=100$. The
parameter $b$ makes the perturbations decay slowly.}
\end{center}
\label{fig4}
\end{figure}
\begin{figure}[H]
\begin{center}
\includegraphics[width=17cm]{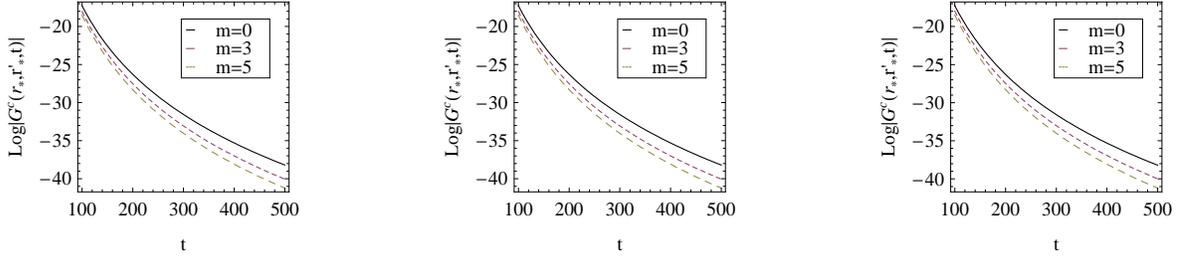}\caption{
Graphs of $\ln G^C(r_*, r'_*; t)$ (the left $s=0$, the middle
$s=1$, the right $s=2$) versus $t$ for the change of $m$, with
$l=5$ and $b=0.9$. Here, $M=1$, $r_*=10$ and $r'_*=100$. The
parameter $m$ makes the perturbations decay quickly.}
\end{center}
\label{fig5}
\end{figure}
\begin{figure}[H]
\begin{center}
\includegraphics[width=18cm]{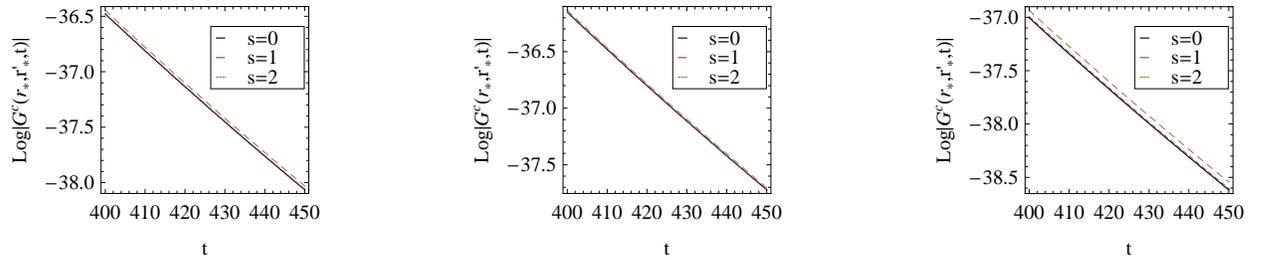} \caption{
Graphs of $\ln G^C(r_*, r'_*; t)$ (the left $b=0.9$, the middle
$b=0.93$, the right $b=0.95$) versus $t$ for the change of $s$,
with $l=5$ and $m=3$. Here, $M=1$, $r_*=10$ and $r'_*=100$. The
spin $s$ makes the perturbations decay slowly.}
\end{center}
\label{fig6}
\end{figure}

\subsection{Late-time behavior of the massive perturbational field}

Here we just consider the late-time tails of the massive scalar
field because in modern physics the electromagnetic and
gravitational fields are massless. As in the massless case, for
the massive scalar field we will also evaluate the contribution of
$G^c(r_*,r'_*;t)$ to the late-time tail. The only difference in
the latter is the integral of the Green's function $G(r_*,r'_*;t)$
around the branch cut performs in the interval $-\mu \leq \omega
\leq \mu$ \cite{Koyama} ( $\mu$ is the mass of the scalar
particle) rather than along the total negative imaginary
$\omega$-axis. Assuming that both the observer and the initial
data are situated far away from the black hole so that $r\gg M$,
we can expand the wave equation (\ref{e2}) for the massive scalar
fields as a power series in $M/r$ and obtain ( neglecting terms of
the order $O(\frac{M^2}{r^2})$ and the higher orders)
\begin{eqnarray}
\left[\frac{d^2}{dr^2}+\omega^2-\mu^2+\frac{4M'\omega^2-2M'\mu^2}{r}
-\frac{\rho^2-\frac{1}{4}}{r^2}\right]\zeta(r,\omega)=0\label{e3}.
\end{eqnarray}
For the massive scalar field in this background, the intermediate
late-time tail (which is the tail in the range $\frac{M}{b}\ll
r\ll t\ll bM/(\mu M)^2$) has the behavior
\begin{eqnarray}
G^C(r_*,r'_*;t)&=&\frac{(1+e^{(2\rho+1)i\pi})}{\pi \rho
2^{-3\rho-2}}\frac{\Gamma (-2\rho)\Gamma (\frac{1}{2}+\rho)\Gamma
(1+\rho)\mu^{\rho}}{\Gamma (2\rho)\Gamma (\frac{1}{2}-\rho)}\nonumber \\
&&(r'_*r_*)^{\frac{1}{2}+\rho}t^{-\rho-1}\cos{[\mu t-\frac{\pi
(\rho+1)}{2}]}. \label{G4}
\end{eqnarray}
Comparing equation (\ref{G4}) with equation (\ref{mlg}), we find
that the power-law tail of the massive scalar field at a fixed
radius in the intermediate late-time depends not only on the
structure parameter $b$ of the background metric, but also on the
multiple number $l$ and the angular quantum number $m$ of the wave
modes. Moreover, the intermediate late-time behavior of the
massive scalar field is dominated by an oscillatory inverse
power-law tail which decays slower than that of the massless case.
\begin{figure}[ht]
\begin{center}
\includegraphics[width=6cm]{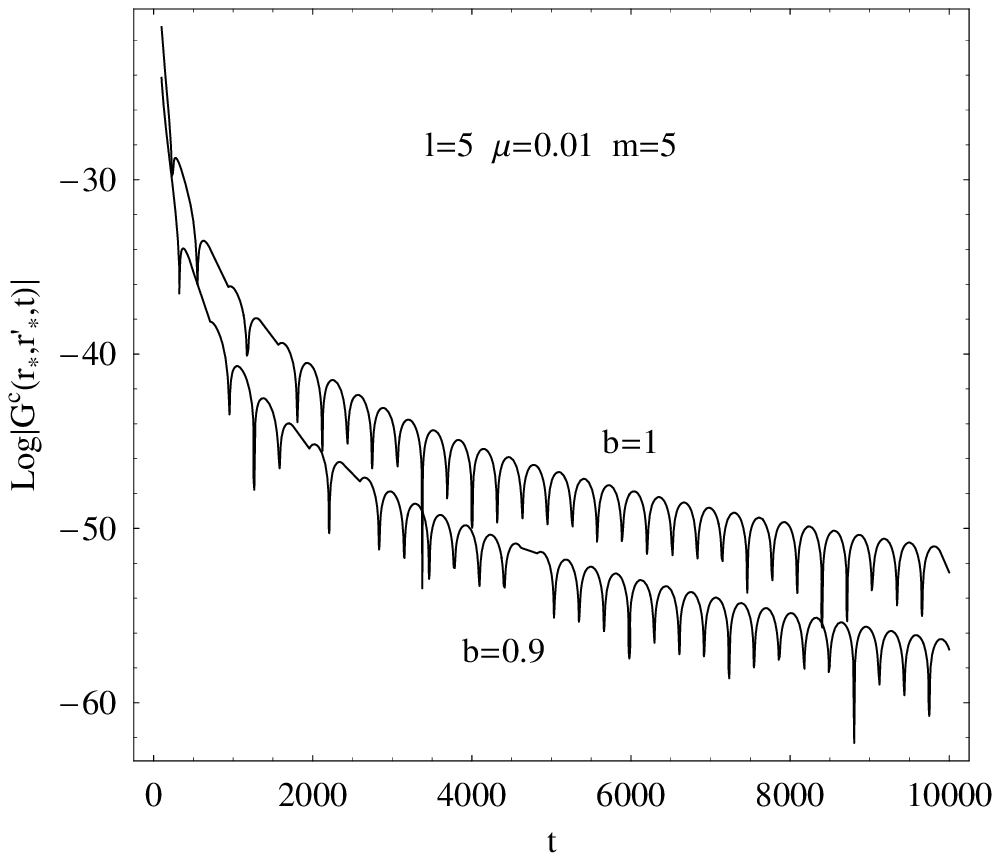}\ \ \ \ \ \ \includegraphics[width=6cm]{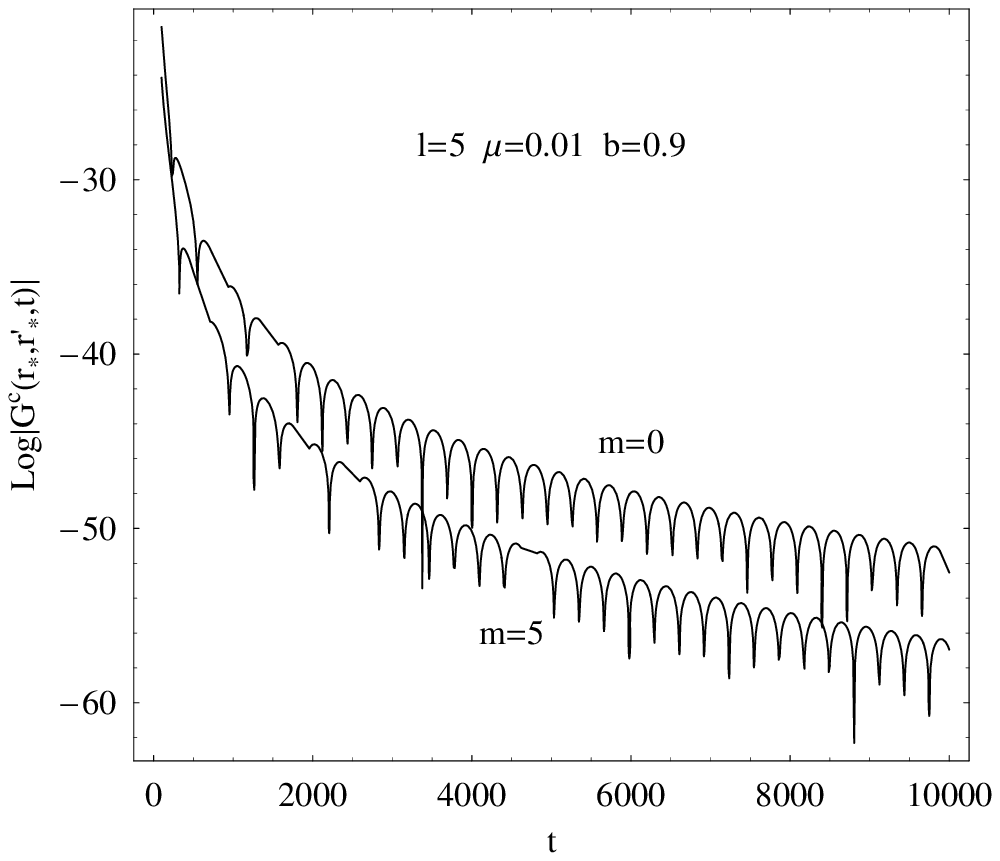}
\caption{The left and the right graphs show $\ln|G^C(r_*, r'_*;
t)|$ versus $t$ for different $b$ and $m$, respectively. Here,
$M=1$, $r_*=10$ and $r'_*=100$.}
\end{center}
\label{fig7}
\end{figure}
Furthermore, figure 7 also tells us that with the increase of $b$,
the massive scalar field dies out slower. For the case $b\neq 1$,
the perturbation field decays faster when $m$ increases.

We also obtain directly that the asymptotic late-time tail (which
is caused by a resonance backscattering of spacetime curvature at
very late-time $\mu t\gg b^2/(\mu M)^2$) can be described by
\begin{eqnarray}
G^C(r_*,r'_*;t)\simeq
\frac{\mu}{2\sqrt{3}}(2\pi)^{-5/6}(M\mu/b)^{1/3}(\mu
t)^{-5/6}\times \sin \{\mu t-\frac{3}{2}(2\pi M/b)^{2/3}(\mu
t)^{1/3}-\varphi_s(\omega_0)-\frac{\pi}{4}\},
\end{eqnarray}
where $\omega_0\sim \mu \sqrt{1-[2\pi M/(t b)]^{2/3}}$ is the
frequency of wave at the saddle-point. This equation tells us that
the decay rate of the asymptotic late-time tail of the massive
scalar field is still in the form $t^{-5/6}$, which is the same as
that in the usual black hole spacetimes and can be regarded as a
quite general feature for the late-time decay of massive scalar
field. Moreover, the oscillation has the period $2\pi/\mu$ which
is modulated by two types of long-term phase shift. The first term
$2\pi(M¦Ì/b)^{2/3}(\mu t)^{1/3}$ represents a monotonously
increasing phase shift and the second one $\varphi_s(\omega_0)$
denotes a periodic phase shift. Both of them depend on the
parameter $b$, which means that in this background the phase
shifts also depend on the the mass per length of the string.

\section{Conclusions and discussions}

The metric of the Schwarzschild black hole with a straight string passing through it is locally identical
to the Schwarzschild metric, so that the presence of the string will not imprint in the motion of test particles since locally
the geodesics of test particles will not be changed.

In this paper we have studied  QNMs and the late-time tail
behaviors of scalar, electromagnetic and gravitational
perturbations in the Schwarzschild black hole spacetime pierced by
a cosmic string. We have observed physical signatures by the
presence of the cosmic string. With the increase of the mass per
length of the string, we observed that both the real part and the
absolute value of the imaginary part of quasinormal frequencies
decrease. The print of the cosmic string in the QNMs can be read
if we compare with that of the usual Schwarzschild black hole.
Furthermore we observed that with a cosmic string passing through
the Schwarzschild black hole background, quasinormal frequencies
depend also on the angular quantum number $m$, which has not been
observed in the usual spherical black hole background before. With
the increase of the angular quantum number $m$,  both the real
parts and the magnitude of the imaginary parts of the QNM
frequencies for electromagnetic and gravitational fields increase,
however for the scalar perturbation, the real part increases while
the magnitude of the imaginary part decreases. Compared with other
perturbations, the gravitational perturbation can last longer
which could be more interesting to observation.

We have extended our study to the late-time tail behaviors of
scalar, electromagnetic and gravitational perturbations in the
Schwarzschild black hole pierced by a cosmic string. We also
observed the physical prints in terms of the cosmic string.
Although the decay rate for the asymptotic late-time tail of the
massive scalar field is still $t^{-5/6}$, which looks the same as
that in the usual black hole cases, the influence due to the
presence of the cosmic string appears in the late-time tail in the
massless perturbations and the intermediate late-time tail for the
massive scalar field. We found that the damping exponents of the
late-time tail for the massless perturbations and the intermediate
late-time tail for the massive scalar field increase with the
increase of the linear mass density of the cosmic string $\rho_s$
and the angular quantum numbers $m$. Moreover, we find that due to
the presence of linear mass density, the decay rates of the
massless perturbations also depend on the spin $s$ of fields.
Compared with other perturbational fields, the gravitational field
decays slower at timelike infinity in this background.

It would be of great interest to get more insights in the
observational signatures of the cosmic string. This would depend a
lot on the preciseness of the gravitational wave observation in
the future if we expect the QNM or the late time tail to tell us
the its signature. There are some important questions waiting to
be addressed, such as what the constraints on the mass density of
the string are, how close to one $b$ has to be observationally
consistent etc. Answers to these interesting questions are called
for.

\begin{acknowledgments}

This work was partially supported by NNSF of China, Ministry of
Education of China and Shanghai Educational Commission. S. B. Chen's work was partially supported  by
the National Basic Research Program of China under Grant No.
2003CB716300 and the Scientific Research Fund of Hunan Normal
University under Grant No.22040639.
\end{acknowledgments}


\begin{thebibliography}{99}
\bibitem{1} H. P. Nollert, Class. Quant. Grav.{\bf 16}, R159 (1999).

\bibitem{2} K. D. Kokkotas and B. G. Schmidt, Living Rev.
Rel. {\bf 2}, 2 (1999).

\bibitem{3} B. Wang, {\it Braz. J. Phys.} \textbf{35},
1029 (2005).

\bibitem{4} G. T. Horowitz and V. E. Hubeny, Phys.
Rev. D {\bf 62}, 024027 (2000).

\bibitem{5} B. Wang, C.Y. Lin, and E. Abdalla, Phys. Lett. B {\bf 481}, 79 (2000);
B. Wang, C. Molina, and E. Abdalla, Phys. Rev. D {\bf 63}, 084001
(2001); J.M. Zhu, B. Wang, and E. Abdalla, Phys. Rev. D {\bf 63},
124004 (2001).

\bibitem{6} V. Cardoso and J.P.S. Lemos, Phys. Rev. D {\bf 63}, 124015 (2001); V. Cardoso and J.P.S. Lemos, Phys.
Rev. D {\bf 64}, 084017(2001); E. Berti and K. D. Kokkotas, Phys.
Rev. D {\bf 67}, 064020 (2003); V. Cardoso and J. P. S. Lemos,
Class. Quant. Grav. {\bf 18}, 5257 (2001);E. Winstanley, Phys.
Rev. D {\bf 64}, 104010 (2001); J. Crisstomo, S. Lepe and J.
Saavedra, Class. Quant. Grav. {\bf 21}, 2801-2810 (2004) ; S.
Lepe, F. Mendez, J. Saavedra, L. Vergara, Class. Quant. Grav. {\bf
20} 2417-2428 (2003).

\bibitem{7} D. Birmingham, I. Sachs, S. N. Solodukhin, Phys. Rev. Lett.{\bf 88}
(2002) 151301; D. Birmingham, Phys. Rev. D {\bf 64}, 064024
(2001).

\bibitem{8} B. Wang, E. Abdalla and R. B. Mann, Phys. Rev. D {\bf 65}, 084006 (2002);
J. S. F. Chan and R. B. Mann, Phys. Rev. D {\bf 59}, 064025
(1999).

\bibitem{9} S. Musiri, G. Siopsis, Phys. Lett. B {\bf 576}, 309-313 (2003);
R. Aros, C. Martinez, R. Troncoso, J. Zanelli, Phys. Rev. D {\bf
67}, 044014 (2003); A. Nunez, A. O. Starinets, Phys. Rev. D {\bf
67}, 124013 (2003).

\bibitem{10} E. Abdalla, B. Wang, A. Lima-Santos and W. G. Qiu, Phys. Lett. B  {\bf 538},
435 (2002); E. Abdalla, K. H. Castello-Branco and A. Lima-Santos,
Phys. Rev. D {\bf 66}, 104018 (2002).

\bibitem{11} S. Hod, Phys. Rev. Lett. {\bf  81}, 4293 (1998);
A. Corichi, Phys. Rev. D  {\bf  67}, 087502 (2003); L. Motl, Adv.
Theor. Math. Phys.{\bf 6}, 1135-1162 (2003); L. Motl and A.
Neitzke, Adv. Theor. Math. Phys. {\bf 7}, 307-330 (2003); A.
Maassen van den Brink, J. Math. Phys. {\bf 45}, 327 (2004) ; O.
Dreyer, Phys. Rev. Lett. {\bf 90}, 08130 (2003); G. Kunstatter,
Phys. Rev. Lett. {\bf 90}, 161301 (2003); N. Andersson and C. J.
Howls, Class. Quant. Grav. {\bf 21}, 1623-1642 (2004); V. Cardoso,
J. Natario and R. Schiappa, J. Math. Phys.{\bf 45}, 4698-4713
(2004).

\bibitem{12} V. Cardoso and J. P. S. Lemos, Phys. Rev. D  {\bf 67}, 084020 (2003);
 K. H. C. Castello-Branco and E. Abdalla, gr-qc/0309090;
 Jose Natario and Ricardo Schiappa, Adv. Theor. Math. Phys. {\bf 8},
 1001-1131 (2004).

\bibitem{13} B. Wang, C. Y. Lin and C. Molina, Phys. Rev. D {\bf 70},
064025 (2004).

\bibitem{aa} J. Natario, R. Schiappa, Adv.Theor.Math.Phys. 8 (2004)
1001.

\bibitem{14} B. Linet, gr-qc/9904044, Class. Quant. Grav. {\bf 16}, 2947 (1999).

\bibitem{15} M. Aryal, L. H. Ford and A. Vilenkin, Phys. Rev. D {\bf 34}, 2263 (1986).

\bibitem{Wheeler1} R. Ruffini and J. A. Wheeler, Phys. Today {\bf 24(1)}, 30 (1971);
 C. W. Misner, K. S. Thorne, and J. A.
Wheeler, \textit{gravitation} (Freeman, San Francisco, 1973).

\bibitem{Price1}R. H. Price, Phys. Rev. D {\bf 5}, 2419 (1972).

\bibitem{Hod2}S. Hod and T. Piran,
Phys. Rev. D {\bf 58}, 024017 (1998); L. Barack,
 Phys. Rev. D {\bf 61} 024026 (2000); L. M. Burko and G. Khanna, Phys. Rev. D {\bf 67},
 081502 (2003); E. S. C. Ching, P. T. Leung, W. M. Suen, and K. Young, Phys. Rev.
D {\bf 52}, 2118 (1995); P. R. Brady, S. Droz, and S. M. Morsink
 Phys. Rev. D {\bf 58} 084034 (1998); C. Gundlach, R. H. Price, and J. Pullin,
 Phys. Rev. D {\bf 49}, 883 (1994); L. M. Burko and G. Khanna,
 Phys. Rev. D {\bf 70}, 044018 (2004); V. Cardoso, S. Yoshida, O. J. C. Dias,
 and J. P. S. Lemos, Phys. Rev. D {\bf 68}, 061503(R) (2003).

\bibitem{Koyama}H. Koyama and A. Tomimatsu,
Phys. Rev. D {\bf 63}, 064032 (2001); Phys. Rev. D {\bf 64},
044014 (2001); R. Moderski and M. Rogatko, Phys. Rev. D {\bf 64},
044024 (2001); Phys. Rev. D {\bf 63}, 084014 (2001); Phys. Rev. D
{\bf 72}, 044027 (2005); S. Hod and T. Piran, Phys. Rev. D {\bf
58}, 044018 (1998).

\bibitem{Hod3}S. Hod,
Phys. Rev. D {\bf 58}, 104022 (1998); L. Barack and A. Ori, Phys.
Rev. Lett. {\bf 82}, 4388 (1999); W. krivan, Phys. Rev. D {\bf
60}, 101501(R) (1999); Q. Y. Pan and J. L. Jing, Chin. Phys. Lett.
{\bf 21}, 1873 (2004).

\bibitem{21} S. A. Teukolsky, Phys. Rev. Lett. {\bf 29}, 1114 (1972); Astrophys. J.
{\bf 185},635 (1973); S. A. Teukolsky and W. H. Press, Astrophys.
J.{\bf 193}, 443 (1974).

\bibitem{22} G. F. Torres del Castillo, J. Math. Phys. {\bf 29}, 2078 (1988);
 J. Math. Phys. {\bf 30}, 446 (1989); U. Khanal, Phys. Rev. D {\bf
 28}, 1291 (1983).

\bibitem{23} R. A. Breuer, M. P. Ryan and  S. Waller,
Proc. R. Soc. Lond. A. {\bf 358}, 71-86 (1977); E. Newman and R.
Penrose, J. Math. Phys. {\bf 3}, 566 (1966); J. N. Goldberg,
\textit{etl}, J. Math. Phys. {\bf 8}, 2155 (1967).

\bibitem{wkb} B. F. Schutz and C. M. Will,  Astrophys. J. Lett. Ed.
{\bf 291}, L33 (1985);  S. Iyer  and C. M. Will, Phys. Rev. D {\bf
35}, 3621 (1987); S. Iyer, Phys. Rev. D {\bf 35}, 3632 (1987).

\bibitem{Cardoso2}E. Berti and V. Cardoso, Phys. Rev. D {\bf 74},
104020 (2006).

\bibitem{Leaver}E. W. Leaver, Proc. R. Soc. Lond. {\bf A 402} 285 (1985)
; Phys. Rev. D {\bf 34} 384 (1986).


\end{thebibliography}
\end{document}